\documentclass[sigconf]{acmart}
\usepackage{makecell}
\usepackage{algorithm}
\usepackage{algorithmic}
\AtBeginDocument{%
  }

\setcopyright{none}
\copyrightyear{2018}
\acmYear{2018}
\acmDOI{XXXXXXX.XXXXXXX}
\acmConference[DAC '26]{Design Automation Conference}{June 25-29, 2026}{Long Beach, CA, USA}
\acmISBN{978-1-4503-XXXX-X/2018/06}

\renewcommand\footnotetextcopyrightpermission[1]{}

\begin{document}

\title{IMMSched: Interruptible Multi-DNN Scheduling via Parallel Multi-Particle Optimizing Subgraph Isomorphism}

\author{Boran Zhao$^*$, Hetian Liu$^*$, Zihang Yuan, Yanbin Hu, Wenzhe Zhao, Tian Xia, Pengju Ren$^\dagger$}
\affiliation{
  \institution{Xi'an Jiaotong University}
  \city{Xi'an}
  \state{Shaanxi}
  \country{China}
}
\thanks{*Both authors contributed equally to this research.}
\thanks{$^\dagger$ Corresponding author (pengjuren@xjtu.edu.cn)}

\renewcommand{\shortauthors}{Zhao et al.}

\begin{abstract}
The growing demand for multi-DNN workloads with unpredictable task arrival times has highlighted the need for interruptible scheduling on edge accelerators. However, existing preemptive frameworks typically assume known task arrival times and rely on CPU-based offline scheduling, which incurs heavy runtime overhead and struggles to handle unpredictable task arrivals. Even worse, prior studies have shown that multi-DNN scheduling requires solving an NP-hard subgraph isomorphism problem on large directed acyclic graphs within limited time, which is extremely challenging.

To tackle this, we propose IMMSched, a parallel subgraph isomorphism method that combines Multi-Particle Optimization with the Ullmann algorithm based on a probabilistic continuous-relaxation scheme, eliminating the serial data dependencies of previous works. Finally, a quantized scheduling scheme and a global controller in the hardware architecture further combine multi-particle results for consensus-guided exploration. Evaluations demonstrate that IMMSched achieves orders-of-magnitude reductions in scheduling latency and energy consumption, enabling real-time execution of unpredictable DNN tasks on edge accelerators.
\end{abstract}





\maketitle

\section{Introduction}

\begin{figure}[t]
    \centering
    \includegraphics[width=1.0\columnwidth]{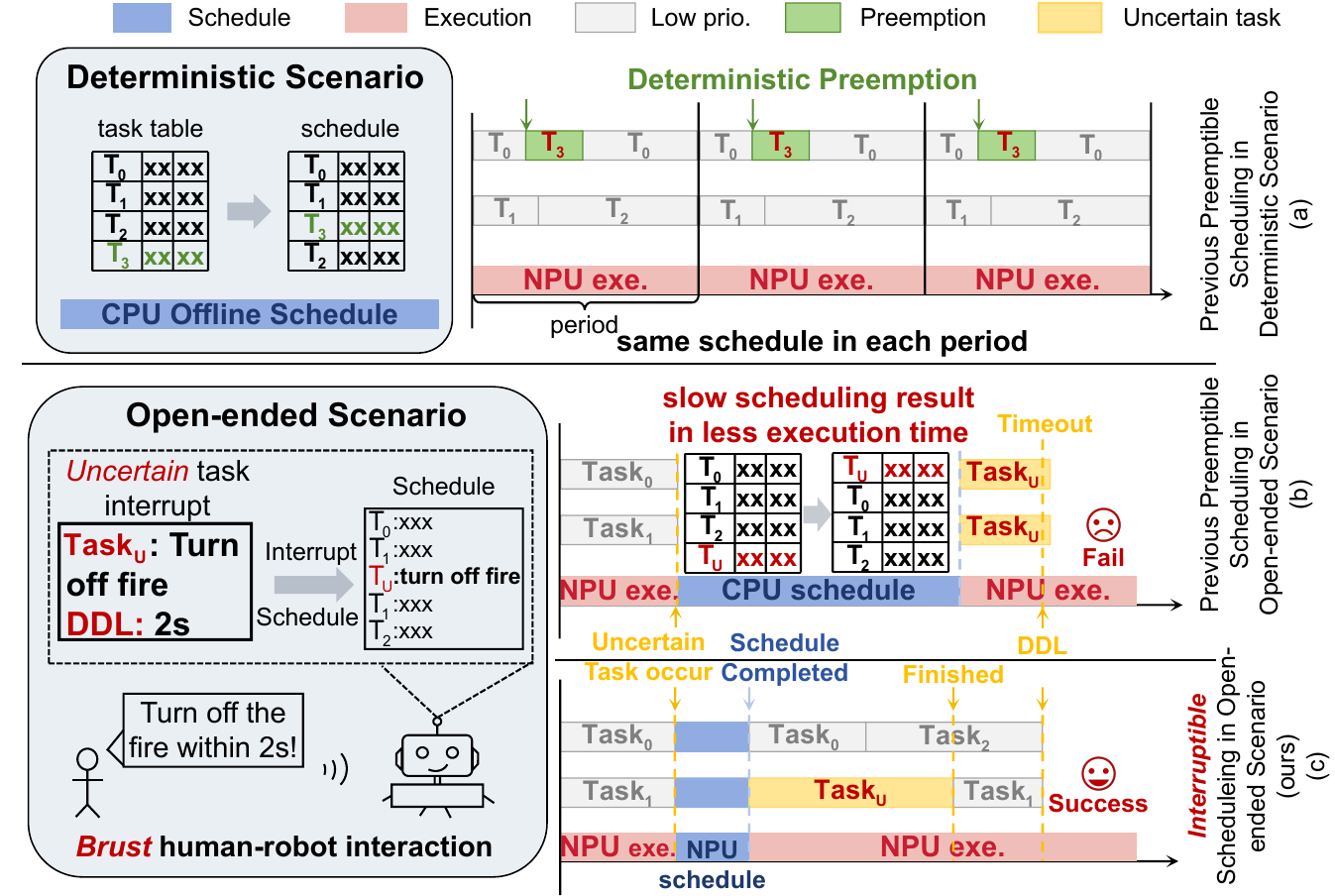}
    \caption{(a) Previous methods target deterministic scenarios, performed offline using CPU-based serial strategies, with tasks executed periodically on the DNN accelerator (i.e., NPU).
(b) In open-ended scenarios, previous methods require significant time to schedule Uncertain tasks, reducing execution time and often causing timeouts.
(c) The proposed interruptible scheduling method leverages the DNN accelerator to parallelize scheduling, ensuring real-time execution of uncertain tasks.}
    \label{fig:previousvsours} 
\end{figure}

The widespread deployment of edge-side multi-DNN workloads in applications such as autonomous driving~\cite{autonomous_driving1}, augmented and virtual reality (AR/VR)~\cite{ARVR} has brought increasing attention to preemptive multi-DNN scheduling architectures (e.g., \cite{HDA,magma}), as they can meet the stringent latency requirements of urgent tasks in some edge scenarios. However, as shown in Table~\ref{tab_sched_framework}, existing preemptive scheduling frameworks~\cite{CD-MSA,hasp,moca,planaria,prema} are primarily designed for scenarios where the trigger time of urgent tasks is known a priori (i.e., tasks are initiated at predetermined or predictable times) and scheduling is typically performed offline using CPU-based serial strategies, as shown in Figure~\ref{fig:previousvsours}.(a).

In contrast, in many real-world multi-DNN scenarios, the execution environment is open-ended, and the arrival of urgent tasks is inherently unpredictable (e.g., in human-robot interaction, robots cannot anticipate spontaneous user commands~\cite{robotic1}, and in autonomous driving, unexpected events such as road hazards may occur without prior warning~\cite{automous_driving_unexcepted1}). These scenarios demand not only timely execution of the urgent task within tight deadlines (DDLs), but also on-the-fly scheduling immediately after task arrival.

\begin{figure}[t]
    \centering
    \includegraphics[width=1.0\columnwidth]{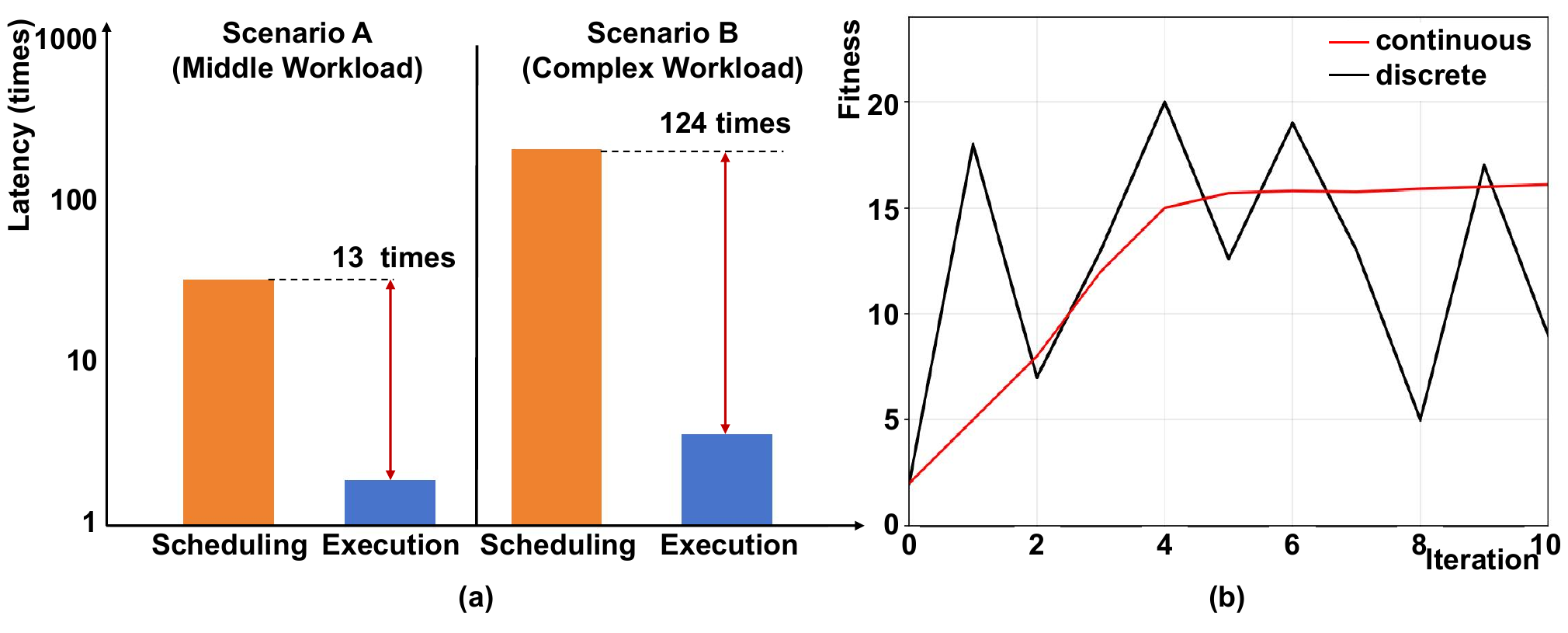}
    \caption{(a) Comparison of execution time and scheduling time on the Cloud platform using MoCA. Scenario A uses the middle workload (i.e., UNet), and Scenario B uses the complex workload (i.e., Qwen). (b) Improved search stability of Particle Swarm Optimization after applying the continuous relaxation mechanism.} 
    \label{fig:schedulevsexe_continvsdisc} 
\end{figure}

Unfortunately, our profiling reveals that traditional CPU-serialized preemptive scheduling incurs significant runtime overhead—often several orders of magnitude larger than the task execution time itself, as shown in Figure~\ref{fig:schedulevsexe_continvsdisc}.(a). As a result, such designs fail to meet the real-time constraints of unpredictable task arrival scenarios, as shown in Figure~\ref{fig:previousvsours}.(b). Therefore, how to efficiently support urgent task scheduling under unpredictable triggers in edge-side multi-DNN environments remains an open and underexplored problem.

\begin{table}[t]
\centering
\caption{Comparison of different scheduling frameworks.}
\label{tab_sched_framework}
\begin{tabular}{ccc c c}
\toprule
 & \makecell{Scheduling\\strategy} & \makecell{Preemptive\\Scheduling} & \makecell{Interruptible\\Scheduling} \\
\midrule
PREMA~\cite{prema} & LTS & \checkmark & $\times$ \\
Planaria~\cite{planaria} & LTS & \checkmark & $\times$ \\
MoCA~\cite{moca} & LTS & \checkmark & $\times$ \\
CD-MSA~\cite{CD-MSA} & LTS & \checkmark & $\times$ \\
HASP~\cite{hasp} & TSS & $\times$ & $\times$ \\
IosSched~\cite{isosched} & TSS & \checkmark & $\times$ \\
IMMSched (Ours) & TSS & \checkmark & \checkmark \\
\bottomrule
\end{tabular}
\end{table}


Meanwhile, prior studies (e.g., IsoSched~\cite{isosched}) have pointed out that performing efficient multi-DNN scheduling on advanced accelerator architectures inherently under cascaded layers pattern (e.g., TSS (Tile Spatial Scheduling)) requires solving increasingly complex subgraph isomorphism problems on large directed acyclic graphs (DAGs). Since the subgraph isomorphism problem is NP-hard~\cite{iso_nphard}, the computational burden becomes prohibitive. Even worse, edge-side systems are often battery-powered and cannot afford additional hardware resources for accelerating subgraph matching. Consequently, efficiently scheduling uncertain DNN tasks under such resource and timing constraints remains a highly challenging problem.


To address the above challenge, we propose IMMSched, an interruptible preemptive scheduling framework for multi-DNN execution. IMMSched leverages the fixed-point matrix and vector computation units already integrated in DNN accelerators to perform rapid subgraph matching, thereby enabling efficient scheduling of unpredictable tasks with negligible additional hardware circuit. Specifically, to exploit the multi-engine architecture of modern accelerators, we propose a parallel multi-particle optimizing (i.e., Particle Swarm Optimization~\cite{PSO}) scheme, where different particles are mapped onto distinct engines to parallelize the Ullmann-based subgraph matching process. However, the discrete nature of the original Ullmann algorithm~\cite{ullmann} often leads to unstable particle search dynamics (as shown in Figure~\ref{fig:schedulevsexe_continvsdisc}.(b)). To mitigate this, we introduce a continuous relaxation mechanism that transforms the discrete matching problem into a continuous optimization process, significantly improving convergence stability. Finally, minor architectural adaptations are incorporated into the accelerator to support the proposed interruptible scheduling flow.


The main contributions of this work are as follows:
\begin{itemize}

\item To the best of our knowledge, this is the first work that supports online interruptible multi-DNN scheduling, enabling efficient execution and preemption of urgent DNN tasks with uncertain trigger times.

\item We propose a parallel subgraph matching algorithm that integrates Particle Swarm Optimization (PSO) with the Ullmann algorithm, effectively removing the data dependencies inherent in prior serial subgraph-matching designs. This innovation enables parallelizable Ullmann execution across multi-engine accelerator architectures.

\item To mitigate the search instability caused by coupling discrete Ullmann matching with continuous PSO optimization, we introduce a probabilistic continuous-relaxation mechanism that transforms the discrete matching problem into a continuous optimization process, thereby stabilizing and accelerating convergence.

\item To execute the continuous subgraph matching efficiently on fixed-point DNN accelerators, we develop a quantized scheduling algorithm and perform lightweight hardware adaptations, including a global coordination controller that fuses multi-particle search results to produce a consensus-guided exploration direction, further improving convergence speed and scheduling efficiency.

\end{itemize}

\section{Background and Motivation}



\subsection{Problems of Previous Works}

\begin{figure}
    \centering
    \includegraphics[width=0.85\linewidth]{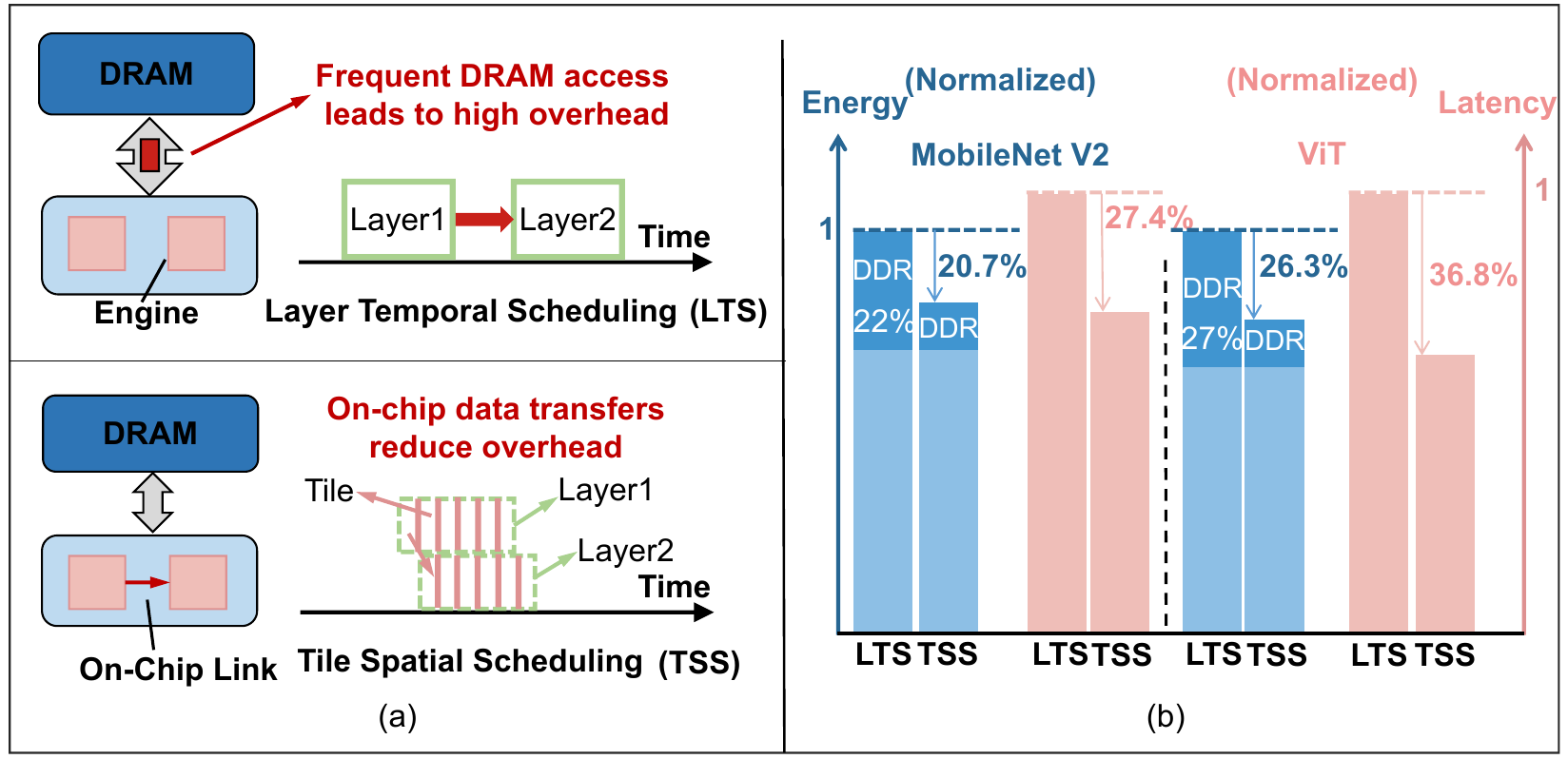}
    \caption{Comparison between Layer Temporal Scheduling (LTS) and Tile Spatial Scheduling (TSS): TSS utilizes on-chip links to avoid the energy and latency overheads associated with DRAM accesses in LTS~\cite{isosched}.}
    \label{fig:LTSvsTTS}
\end{figure}

To address the scheduling problem of multi-DNN tasks, researchers have proposed various methods that can be broadly classified into Layer Temporal Scheduling (LTS) and Tile Spatial Scheduling (TSS), as shown in Table~\ref{tab_sched_framework}. For instance, PREMA~\cite{prema} assigns each task to a single core, while Planaria~\cite{planaria}, MoCA~\cite{moca}, and CD-MSA~\cite{CD-MSA} further refine task execution by introducing parallelism, dynamic memory optimization, and cross-layer overlapping, respectively. However, as shown in Figure~\ref{fig:LTSvsTTS}, all these approaches fundamentally follow the LTS paradigm, where frequent off-chip DRAM accesses between layers lead to high energy consumption and significant performance degradation.

\begin{figure*}[t]
    \centering
    \includegraphics[width=1.7\columnwidth]{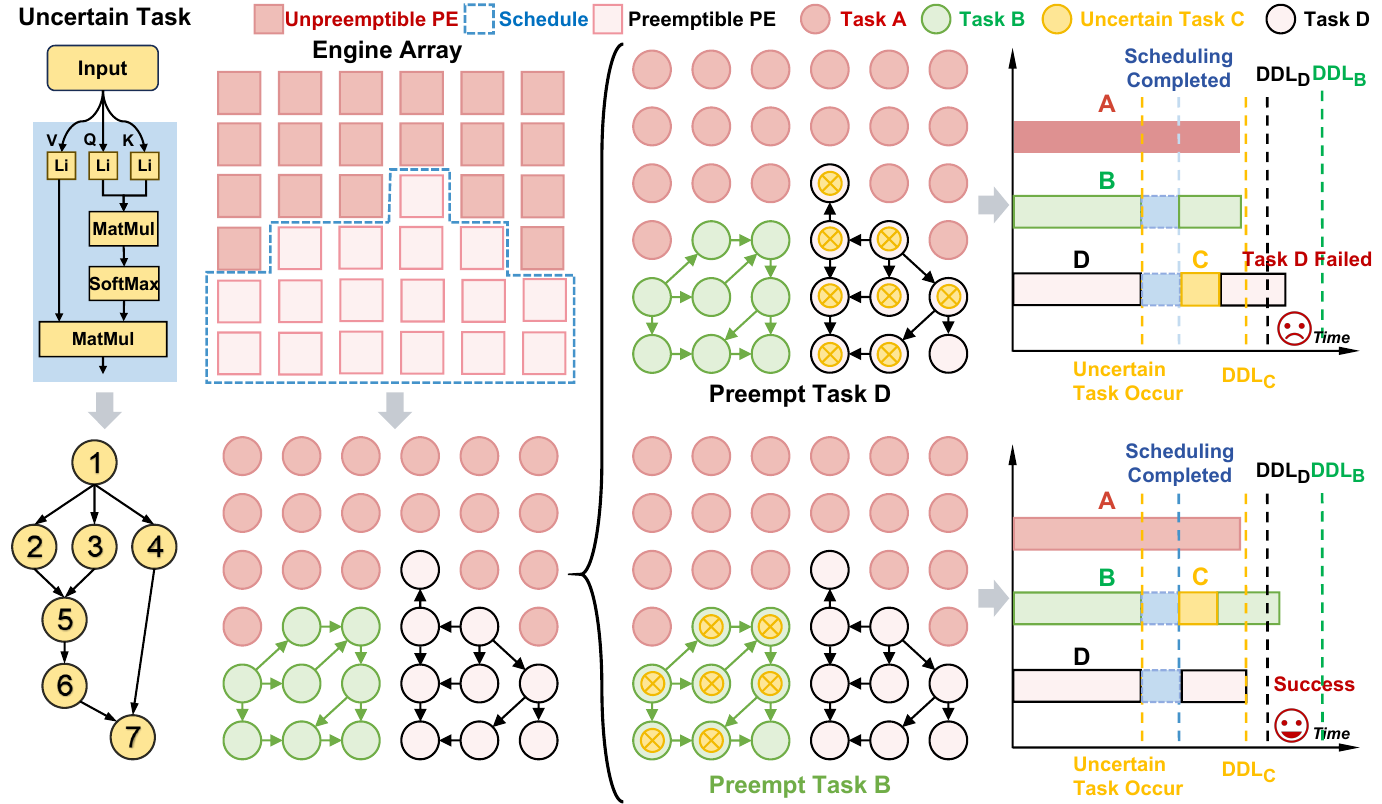}
    \caption{MMSched schedules by preempting engines based on the "single-core preemption ratio," prioritizing low-priority tasks (e.g., Task B and D) according to execution time slack, while not interrupting high-priority tasks (e.g., Task A) by default.}
    \label{fig:subgraphmatching} 
\end{figure*}

To overcome the structural limitations of LTS-based approaches, researchers have proposed several TSS-based algorithms~\cite{hasp,mars,isosched}. Among them, IsoSched~\cite{isosched} is the first to introduce a preemptive scheduling strategy under the TSS paradigm. As shown in Figure~\ref{fig:subgraphmatching}, the key idea of IsoSched is to abstract multiple DNN tasks as a preemptible DAG and formulate the preemptive scheduling of high-priority tasks as a subgraph matching problem, thereby enabling fine-grained preemption within the TSS framework.


However, these algorithms still exhibit two major limitations:
\begin{itemize}
\item Previous preemptive scheduling methods primarily target deterministic tasks. When applied to burst or unexpected tasks, such methods often suffer from long scheduling latency, leaving insufficient time for actual execution. In practice, these highly time-sensitive unexpected tasks are common in dynamic multi-DNN workloads.
\item Although IsoSched, as a TSS-based preemptive scheduler, effectively avoids the high latency overhead caused by frequent DRAM accesses in LTS frameworks, its serial subgraph-matching procedure remains inefficient, making it difficult to respond promptly to high-priority tasks under tight deadlines.
\end{itemize}


\subsection{Inspiration of Subgraph Isomorphism}


Inspired by IsoSched~\cite{isosched}, we similarly abstract the preemptive scheduling problem after interruption as a subgraph matching task. However, subgraph matching is an NP-hard problem, and traditional serial algorithms (e.g., GsPM~\cite{GpSM}, VF2~\cite{vf2}, VF3~\cite{vf3}) either exhibit strong serial dependencies (e.g., backtracking and multi-way joining~\cite{overview_iso}) or involve intensive comparison and selection operations, which are unfriendly to compute-bound, pipelined DNN accelerators. Therefore, it is critical to choose a subgraph matching algorithm that is both parallelizable and compatible with accelerator execution, while carefully addressing its inherent serial dependencies.

\subsection{Challenge of Interruptible Scheduling}

In open-ended scenarios, task trigger is highly unpredictable and interruptible scheduling requires subgraph matching to be performed quickly. However, given the limited on-chip resources, designing dedicated hardware for subgraph matching is not practical. Therefore, designing an interruptible parallel scheduling strategy remains a challenging problem:

\begin{itemize}
    \item The inherent high latency of serial subgraph matching algorithms makes it difficult to meet tight delay constraints, with the key challenge being how to exploit the algorithm's parallelization potential.
    
    \item Conventional subgraph matching involves a discrete, combinatorial search space, making it difficult to create a continuous, differentiable optimization objective. This limits the effectiveness of heuristic methods and challenges stable convergence.
    \item Most optimization algorithms use floating-point arithmetic for high precision, while DNN accelerators are optimized for fixed-point computation, creating a gap in effectively repurposing them for scheduling algorithm acceleration.
    
\end{itemize}

\section{IMMSched}
In this section, we first follow IsoSched~\cite{isosched} to formally model the multi-DNN scheduling problem. We then propose a continuous-relaxation-based scheduling algorithm that integrates multi-particle optimizing and parallel subgraph matching. Finally, we perform algorithm quantization and implement a lightweight accelerator adaptation along with a global controller to enable efficient execution on real hardware.

\subsection{Preliminary of Formalization}


Following IsoSched~\cite{isosched}, we first formulate the multi-DNN scheduling problem using Integer Linear Programming and construct two scheduling tensors:
\[
X \in \{0,1\}^{D\times I\times N\times T\times P}, \quad 
Y \in \{0,1\}^{D\times I\times K\times T\times L}
\]
representing the compute and communication mappings, respectively. 
We then employ the DAG-to-Pipeline method proposed in ReMap~\cite{remap} and the Layer Concatenate-and-Split mechanism from IsoSched~\cite{isosched} to construct a preemptible DAG based on tensors $X$ and $Y$. Finally, a subgraph matching algorithm is executed to identify feasible scheduling strategies.


Unlike IsoSched, our approach adopts a continuous-relaxation modeling technique, which transforms the inherently discrete subgraph matching problem into a continuous optimization problem solvable in differentiable space. Building on this formulation, we integrate a multi-particle optimizing algorithm~\cite{PSO} to parallelize and reconstruct the Ullmann algorithm\cite{ullmann}, proposing a parallel subgraph matching framework capable of efficiently searching feasible mappings in continuous space.


\subsection{Proposed Continuous-Relaxation Modeling}

We first abstract the DNN workload and the preemptible PE array of the accelerator as two DAGs. The workload and accelerator are represented by their respective adjacency matrices, denoted as the query graph $Q \in\{0,1\}^{n \times n}$ and the target graph $G \in \{0,1\}^{m \times m}$, where $n$ and $m$ denote the number of vertices in the query and target graph, respectively.

Taking into account both the in/out-degree relationships of $Q$ and $G$ and the computation type of each vertex (e.g., convolution for compute-intensive tiles, and max-pooling for comparison-intensive tiles), we construct a global compatibility mask matrix $\mathrm{Mask} \in \{0,1\}^{n\times m}$. Each element $\mathrm{mask}_{ij}=1$ indicates that the $i$-th tile of the DNN workload can be mapped onto the $j$-th PE in the accelerator.


To address the instability caused by coupling the discrete Ullmann algorithm with continuous PSO algorithm, we perform a continuous relaxation of the mapping relation based on the global mask matrix. Specifically, we define a continuous relaxation matrix $S \in [0,1]^{n\times m}$, where each row of $S$ sums to~1. The element $s_{ij}$ represents the probability that the $i$-th tile in the DNN workload is mapped to the $j$-th PE in the accelerator. Leveraging the inherent parallelism of the multi-particle optimizing algorithm---in which particles are relatively independent and easy to parallelize---we distribute multiple particles, each corresponding to a continuous relaxed mapping $S$, across the multi-core DNN accelerator for parallel iterative updates. The detailed algorithm is provided in the Section~\ref{subsec:parallel subgrapg matching}.

\begin{algorithm}[!t]
\caption{Ullmann-refined PSO for Subgraph Matching}\label{alg:Ullmann-PSO}
\begin{algorithmic}[1]

\REQUIRE Query adjacency $Q\in\{0,1\}^{n\times n}$, target adjacency $G\in\{0,1\}^{m\times m}$, number of particles $N$, epochs $T$, inner steps $K$
\ENSURE Set of feasible mappings $\mathcal{M}$

\STATE $\mathcal{M} \gets \emptyset$
\STATE $\mathrm{Mask} \gets \textsc{InitCompatibilityMask}(Q,G)$
\FOR{$t=1$ to $T$}
    \STATE $\mathcal{P} \gets \textsc{InitParticles}(N)$
    \STATE $\bar{S} \gets \emptyset$
    \FOR{each $p_i$ $\in$ $\mathcal{P}$}
        \FOR{$k=0$ to $K-1$}
            \STATE $V_i^{(k+1)} \gets \textsc{Velocity}\!\left(V_i^{(k)},\,S_i^{(k)},\,S^*,\,S_{local},\,\bar{S}\right)$
            \STATE $S_i^{(k+1)} \gets \textsc{Position}\!\left(S_i^{(k)},\,V_i^{(k+1)}\right)$
            \STATE $S_i^{(k+1)} \gets S_i^{(k+1)} \odot \mathrm{Mask}$
            \STATE $S_i^{(k+1)} \gets \textsc{RowNormalize}\!\left(S_i^{(k+1)}\right)$
            \IF{$f_i>f_{local}$}
                \STATE $f_{local},\,S_{local}\gets \textsc{Upadte}(f_i,\,S_i)$
            \ENDIF
            \IF{$f_i>f^*$}
                \STATE $f^*,\,S^*\gets \textsc{Upadte}(f_i,\,S_i)$
            \ENDIF  
        \ENDFOR
        \STATE $\widetilde{M}_i \gets \textsc{Projection}\!\left(S_i^{(K)},\,\mathrm{Mask}\right)$
        \STATE $\widehat{M}_i \gets \textsc{UllmannRefine}\!\left(\widetilde{M}_i,\,Q,\,G\right)$
        \STATE $f_i \gets \textsc{EvaluateFitness}\!\left(S_i^{(K)},\,Q,\,G,\,\mathrm{Mask}\right)$
   
        \IF{$\textsc{IsFeasible}\!\left(\widehat{M}_i,\,Q,\,G\right)$}
            \STATE $\mathcal{M} \gets \mathcal{M} \cup \{\widehat{M}_i\}$
            \STATE $\bar{S} \gets \textsc{EliteConsensus}\!\left(\{(S_i^{(K)},f_i)\}_{i=1}^{N}\right)$
        \ENDIF
    \ENDFOR
\ENDFOR
\RETURN $\mathcal{M}$

\end{algorithmic}
\end{algorithm}

\subsection{Proposed Parallel Subgraph Matching Algorithm}
\label{subsec:parallel subgrapg matching}


To eliminate the serial dependencies inherent in traditional subgraph matching algorithms and to avoid excessive comparison and selection operations during problem solving, we take the Ullmann algorithm as the foundation of our design and develop a parallelized extension. The Ullmann algorithm is particularly well-suited for this purpose because its serial dependencies can be decomposed into parallel exploration steps, serving as heuristic guidance for the multi-particle optimizing algorithm to steer the search direction. Moreover, Ullmann inherently performs feasibility verification through matrix multiplication, which naturally aligns with the compute-intensive characteristics of DNN accelerators.


To achieve a parallel reformulation of subgraph matching, we integrate the PSO with the Ullmann algorithm. The PSO enables global exploration through population-based search, effectively avoiding local optima. In addition, the relative independence of each particle makes it inherently parallelizable, which perfectly matches our goal of designing an efficient and scalable parallel scheduling algorithm.



Specifically, as illustrated in Algorithm~\ref{alg:Ullmann-PSO}, each particle maintains a continuously relaxed mapping matrix $S$ and its particle-wise local optimum $S_{\text{local}}$, while the global controller maintains the set of all feasible mappings $\mathcal{M}$, the global best particle $S^{*}$, and a consensus matrix $\bar{S}$, which jointly guide the velocity update term $V$ for each generation. To ensure that each particle remains valid after the update, we apply both the global mask matrix and a normalization constraint to the updated mapping. Meanwhile, based on the structural analysis of the Ullmann algorithm, we employ an edge-preserving metric $\| Q - S G S^{\mathsf{T}} \|^{2}$ to measure the accuracy of the mapping in preserving the adjacency structure of the query graph. 






Finally, the mapping $S$ is projected onto a valid discrete mapping $\widehat{M}_i$, where each vertex in the query graph is mapped to exactly one vertex in the target graph, and each target vertex can be matched by at most one query vertex. The Ullmann algorithm is invoked to verify the feasibility of each candidate mapping by checking whether $\widehat{M}_i G \widehat{M}_i^{\mathsf{T}}$ contains the query graph $Q$.





In addition, the running tasks are classified into different priority levels according to their urgency. When an interrupt is triggered, the IMMScheduler selects low-priority tasks for preemption according to a adaptive "single-core preemption ratio" policy, as shown in Figure~\ref{fig:subgraphmatching}. When subgraph matching succeeds and multiple feasible mappings are found, the IMM Scheduler prioritizes preempting the task with the largest execution-time slack, so as to avoid deadline violations of the original tasks caused by preemption.

\subsection{Algorithm Quantization and Hardware Improvement}




We first perform a static analysis of the value ranges of the key variables in the algorithm. The binary matrices (i.e., global compatibility mask matrix $\mathrm{Mask}$, query matrix $Q$, and target matrix $G$) are represented using $\{0,1\}$, and the relaxed matrix $S$ is uniformly quantized into 8-bit unsigned integers. 

Under this representation, both the binary matrices and the quantized $S$ can be directly mapped to the accelerator’s \texttt{int8}/\texttt{uint8} MAC datapath, with accumulation performed in \texttt{int32}, thereby avoiding the overhead of floating-point computation. 

\begin{figure}[t]
    \centering
    \includegraphics[width=0.8\columnwidth]{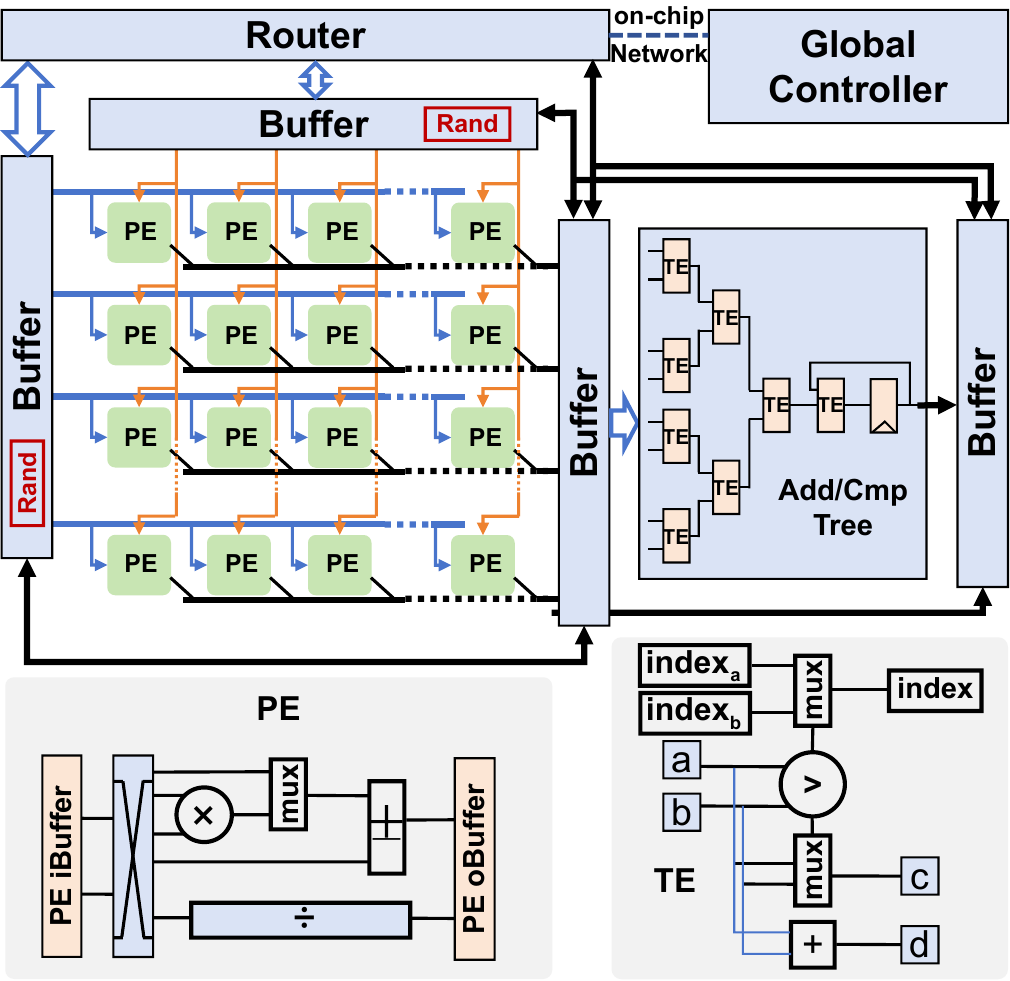}
    \caption{Hardware architecture supporting IMMSched based on typical DNN accelerator.} 
    \label{fig:hw} 
\end{figure}

We analyzed the dataflow of the algorithm and decomposed it into several computation patterns (e.g., element-wise matrix operations and vector summation). According to the characteristics of these patterns, we performed lightweight modifications on a typical DNN accelerator. 
As shown in the lower left part of Figure~\ref{fig:hw}, to element-wise matrix operations, arbiters and selectors were added to the existing PEs to enable different operations. In addition, considering the high hardware cost of division, we replaced divider with a multiplication by a reconfigurable reciprocal value.

As illustrated in Figure~\ref{fig:hw}, to support the vector summation, the tree-based accumulator was redesigned by adding comparators and selectors, enabling the output of the index corresponding to the maximum value within a vector. 
Moreover, we designed a lightweight on-chip global controller to map the parallel subgraph matching algorithm onto the DNN accelerator. The controller is directly connected to the accelerator array through the on-chip network.



\begin{figure*}[t]
    \centering
    \includegraphics[width=1.85\columnwidth]{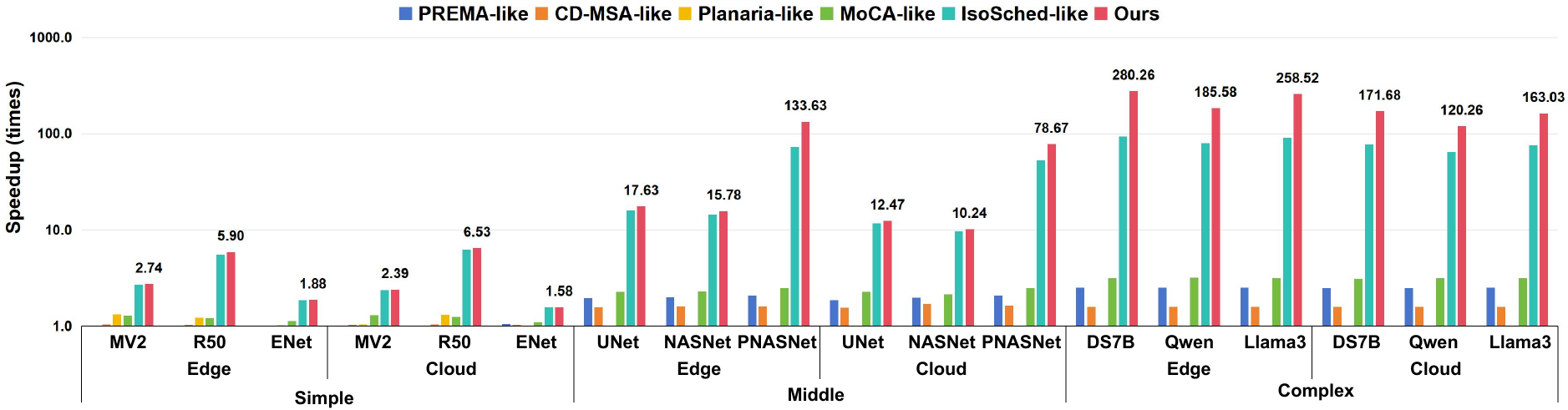}
    \caption{Normalized speedup across various platforms (e.g., Edge and Cloud) and workloads (e.g., Simple, Middle and Complex). } 
    \label{fig:latency} 
\end{figure*}

\begin{table}[t]
\caption{Hardware resource constrains of different platforms}
\centering
\resizebox{0.4\textwidth}{!}{ 
\begin{tabular}{l c c c}
\hline
\textbf{Platform} & \textbf{MACs} & \textbf{Engines} & \textbf{Clock Frequency} \\
\hline
Edge & 64 & $128\times $128  & 700MHz\\
\hline
Cloud & 128 & $128\times $128  & 700MHz\\
\hline
\end{tabular}
}

\label{tab_platforms}

\end{table}

\begin{figure}[t]
    \centering
    \includegraphics[width=0.9\columnwidth]{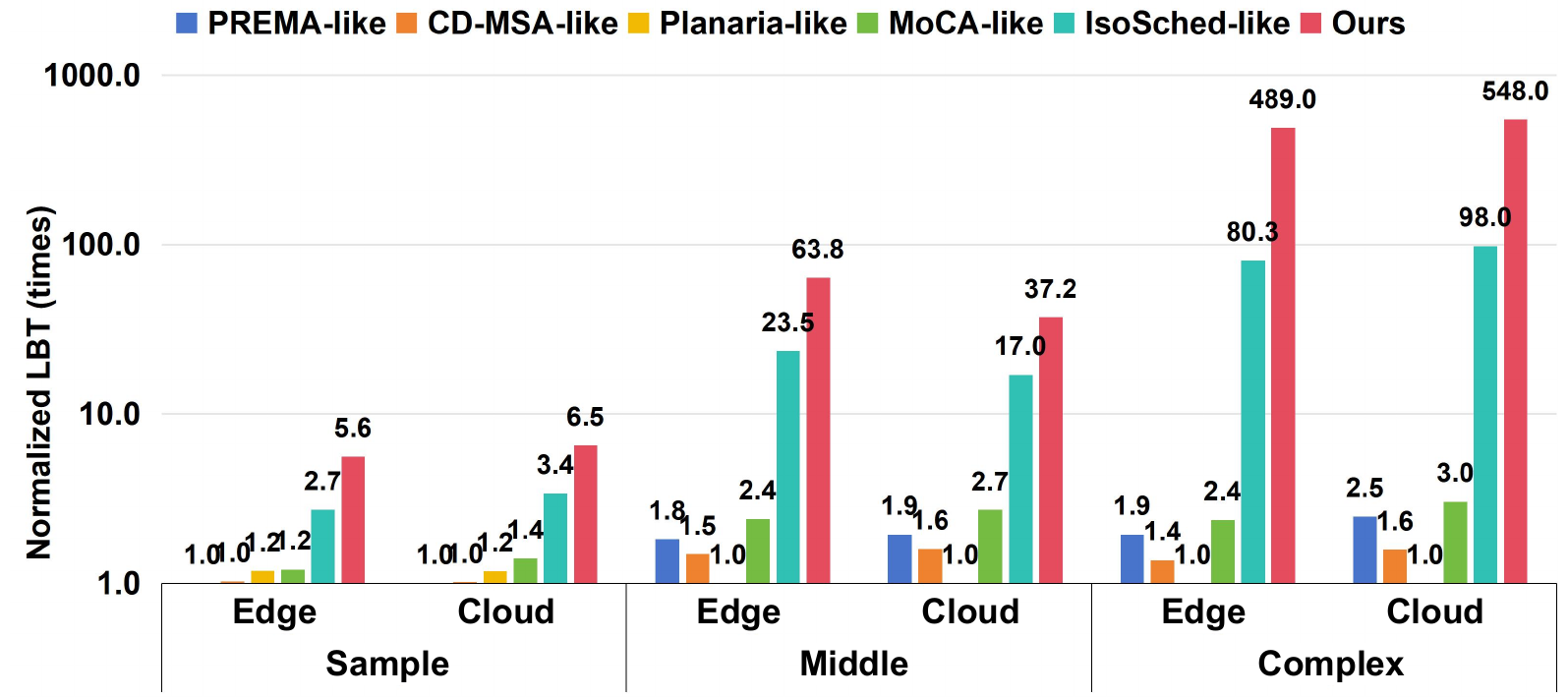}
    \caption{Normalized LBT performance across various platforms and workloads.} 
    \label{fig:LBT} 
\end{figure}

\begin{figure}[t]
    \centering
    \includegraphics[width=0.9\columnwidth]{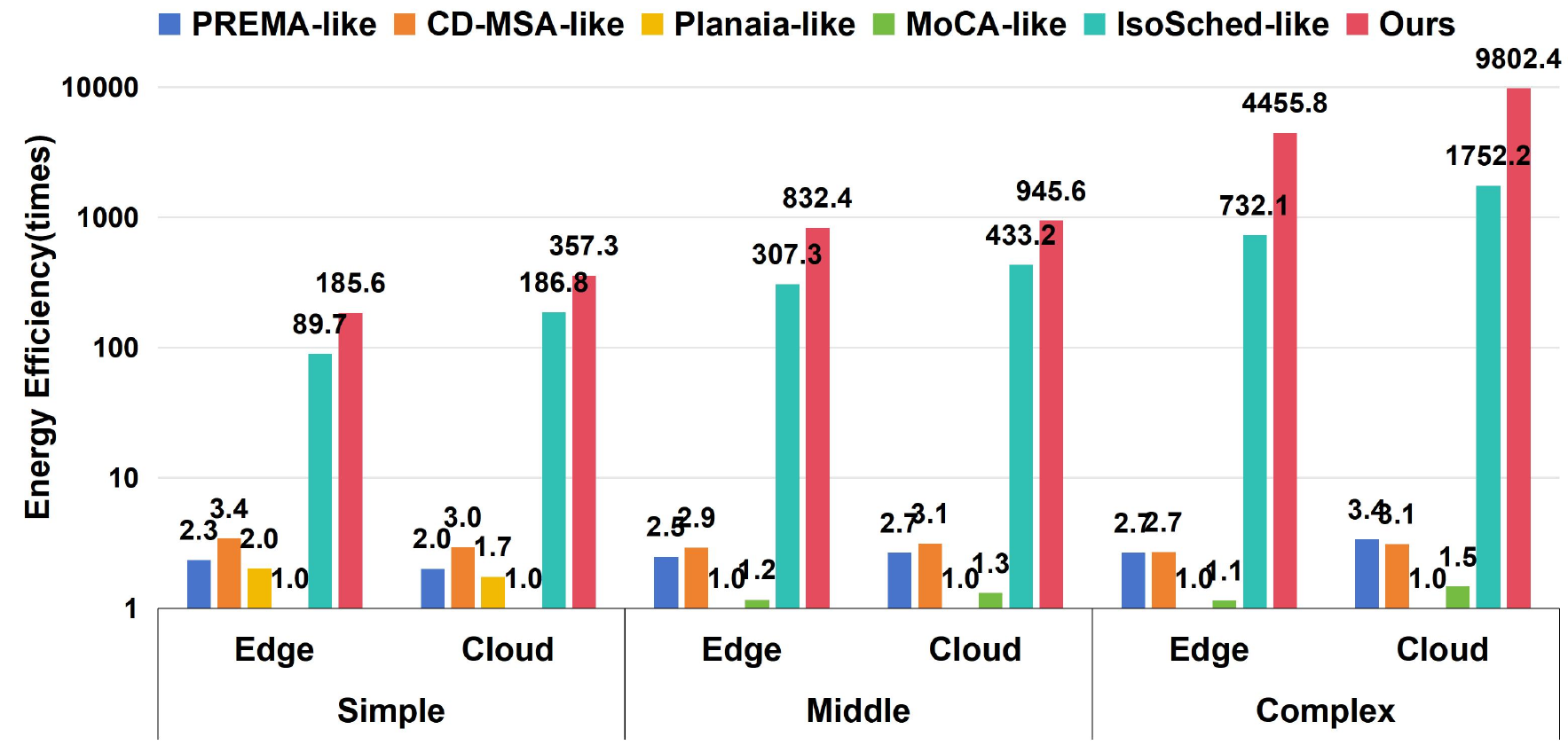}
    \caption{Normalized energy efficiency across various platforms and workloads of different complexities.} 
    \label{fig:energy} 
\end{figure}

\section{Experiment and Analysis}

\subsection{Experimental Setups}

\subsubsection{Hardware Modeling}
Following Planaria~\cite{planaria} and MoCA~\cite{moca}, we implement the engines and NoC in Verilog and synthesize them with Synopsys Design Compiler (T-2022.03-SP5) using the FreePDK 45nm~\cite{FreePDK45} standard cell library to obtain power and area. On-chip SRAM is modeled with CACTI-P~\cite{cacti-p} to report both energy and area. The NoC is modeled using McPAT 1.3~\cite{mcpat}, with a per-hop energy of 0.64 pJ/bit. Following IsoSched~\cite{isosched}, the overall evaluation platform is summarized in Table~\ref{tab_platforms}.



\subsubsection{Workloads}


To evaluate IMMSched's scheduling efficiency under DNN workloads of varying complexity, we define three categories: Simple, Middle, and Complex. As shown in Figure~\ref{fig:latency}, the Simple workload includes MobileNetV2~\cite{Mobilenetv2}, ResNet50~\cite{resnet50}, and UNet~\cite{U-net}, commonly used in AR/VR. The Middle workload includes EfficientNet~\cite{Efficientnet}, NASNet~\cite{NASNet}, and PNASNet~\cite{PNASNet}, typically used in NAS. For the Complex workload, we select DeepSeek-7B~\cite{deepseek}, Qwen-7B~\cite{qwen}, and Llama-3-8B~\cite{Llama-3}, which represent deeper models with higher computational and communication complexity.

%

\subsubsection{Baseline}

IMMSched is designed for open-ended environments, we only select preemptive scheduling algorithms as baselines. Specifically, we choose PREMA~\cite{prema}, CD-MSA~\cite{CD-MSA}, Planaria~\cite{planaria} and MoCA~\cite{moca} as LTS-based baselines, and IsoSched~\cite{isosched} as the TSS-based baseline. 

\subsubsection{Metric}
The comparison is conducted in terms of three key metrics: Speedup, Latency-Bound Throughput (LBT), and Energy efficiency. 
Following PREMA \cite{prema}, Planaria\cite{planaria}, and CD-MSA\cite{CD-MSA}, LBT is defined as the maximum queries-per-second ($1/\lambda$) achieved by the system under a Poisson arrival rate~$\lambda$. LBT measures whether urgent tasks in multi-DNN scheduling can be successfully satisfied. Following IsoSched~\cite{isosched}, Speedup reflects the reduction in task total latency compared to the baseline, where total latency reflects the overall scheduling delay, comprising both scheduling and execution. Energy Efficiency denotes the throughput achieved per unit of energy consumption~\cite{energy}, and evaluates the impact of on-chip link utilization.



\subsection{Comparison with Baselines}

\subsubsection{Speedup}

as shown in Figure~\ref{fig:latency}, across various platforms (i.e., Edge and Cloud) and workloads with different topological complexities (i.e., Simple, Middle, Complex), compared to \emph{LTS-based} approaches including PREMA-like, CD-MSA-like, Planaria-like and MoCA-like, IMMSched achieves average improvements in Speedup by $\times$34.4, $\times$51.4, $\times$81.4, and $\times$27.9, respectively, compared to \emph{TSS-based} approach Isosched, IMMSched achieves average improvements in Speedup by $\times$1.6. These improvements stem from IMMSched performing scheduling on the DNN accelerator and using on-chip data transfers, which greatly reduces computation and communication latency. In addition, quantization lowers bandwidth requirements for data storage and movement, further reducing latency and improving speedup.

\subsubsection{Latency-Bound Throughput}    


as shown in Figure~\ref{fig:LBT}, across various platforms and workloads with different topological complexities, compared to \emph{LTS-based} approaches including PREMA-like, CD-MSA-like, Planaria-like and MoCA-like, IMMSched achieves average improvements in LBT by $\times$89.8, $\times$130.2, $\times$191.4, and $\times$72.7 respectively, compared to \emph{TSS-based} approach Isosched, IMMSched achieves average improvements in LBT by $\times$3.4. The main improvement comes from IMMSched’s use of PSO for parallel subgraph isomorphism search, while the continuous relaxation smooths the search space and accelerates convergence.

\subsubsection{Energy Efficiency}

as shown in Figure~\ref{fig:energy}, across various platforms and workloads with different topological complexities, compared to \emph{LTS-based} approaches including PREMA-like, CD-MSA-like, Planaria-like and MoCA-like, IMMSched achieves average improvements in Energy Efficiency by $\times$918.6, $\times$927.9, $\times$2722.2, and $\times$2092.7, respectively, compared to \emph{TSS-based} approach Isosched, IMMSched achieves average improvements in Energy Efficiency by $\times$3.43. The improvements can be attributed to the TSS paradigm, which uses on-chip data movement to reduce DRAM energy, and to the additional gains from reusing the DNN accelerator for customized computations, further enhancing energy efficiency.






\section{Conclusion}


This paper presents IMMSched, an interrupt-driven multi-DNN scheduling framework designed for open-ended scenarios. IMMSched transforms scheduling into a search for feasible subgraph isomorphisms, where the Ullmann algorithm is redesigned and parallelized using PSO to accelerate the subgraph matching process. A parallel subgraph matching method is developed based on continuous relaxation modeling, enabling compatibility with the continuous optimization mechanism of PSO. Furthermore, lightweight hardware improvements and fixed-point quantization are introduced to enable the scheduling algorithm to run efficiently on practical DNN accelerators. Compared with existing scheduling approaches, IMMSched effectively addresses the urgent task scheduling challenges in open-ended environments, which previous methods fail to handle.


\bibliographystyle{ACM-Reference-Format}
\bibliography{ref}


\begin{thebibliography}{34}


\ifx \showCODEN    \undefined \def \showCODEN     #1{\unskip}     \fi
\ifx \showISBNx    \undefined \def \showISBNx     #1{\unskip}     \fi
\ifx \showISBNxiii \undefined \def \showISBNxiii  #1{\unskip}     \fi
\ifx \showISSN     \undefined \def \showISSN      #1{\unskip}     \fi
\ifx \showLCCN     \undefined \def \showLCCN      #1{\unskip}     \fi
\ifx \shownote     \undefined \def \shownote      #1{#1}          \fi
\ifx \showarticletitle \undefined \def \showarticletitle #1{#1}   \fi
\ifx \showURL      \undefined \def \showURL       {\relax}        \fi
\providecommand\bibfield[2]{#2}
\providecommand\bibinfo[2]{#2}
\providecommand\natexlab[1]{#1}
\providecommand\showeprint[2][]{arXiv:#2}

\bibitem[Bai et~al\mbox{.}(2023)]%
        {qwen}
\bibfield{author}{\bibinfo{person}{Jinze Bai}, \bibinfo{person}{Shuai Bai}, \bibinfo{person}{Yunfei Chu}, \bibinfo{person}{Zeyu Cui}, \bibinfo{person}{Kai Dang}, \bibinfo{person}{Xiaodong Deng}, \bibinfo{person}{Yang Fan}, \bibinfo{person}{Wenbin Ge}, \bibinfo{person}{Yu Han}, \bibinfo{person}{Fei Huang}, {et~al\mbox{.}}} \bibinfo{year}{2023}\natexlab{}.
\newblock \showarticletitle{Qwen technical report}.
\newblock \bibinfo{journal}{\emph{arXiv preprint arXiv:2309.16609}} (\bibinfo{year}{2023}).
\newblock


\bibitem[Bi et~al\mbox{.}(2024)]%
        {deepseek}
\bibfield{author}{\bibinfo{person}{Xiao Bi}, \bibinfo{person}{Deli Chen}, \bibinfo{person}{Guanting Chen}, \bibinfo{person}{Shanhuang Chen}, \bibinfo{person}{Damai Dai}, \bibinfo{person}{Chengqi Deng}, \bibinfo{person}{Honghui Ding}, \bibinfo{person}{Kai Dong}, \bibinfo{person}{Qiushi Du}, \bibinfo{person}{Zhe Fu}, {et~al\mbox{.}}} \bibinfo{year}{2024}\natexlab{}.
\newblock \showarticletitle{Deepseek llm: Scaling open-source language models with longtermism}.
\newblock \bibinfo{journal}{\emph{arXiv preprint arXiv:2401.02954}} (\bibinfo{year}{2024}).
\newblock


\bibitem[Carletti et~al\mbox{.}(2017)]%
        {vf3}
\bibfield{author}{\bibinfo{person}{Vincenzo Carletti}, \bibinfo{person}{Pasquale Foggia}, \bibinfo{person}{Alessia Saggese}, {and} \bibinfo{person}{Mario Vento}.} \bibinfo{year}{2017}\natexlab{}.
\newblock \showarticletitle{Challenging the time complexity of exact subgraph isomorphism for huge and dense graphs with VF3}.
\newblock \bibinfo{journal}{\emph{IEEE transactions on pattern analysis and machine intelligence}} \bibinfo{volume}{40}, \bibinfo{number}{4} (\bibinfo{year}{2017}), \bibinfo{pages}{804--818}.
\newblock


\bibitem[Carletti et~al\mbox{.}(2015)]%
        {vf2}
\bibfield{author}{\bibinfo{person}{Vincenzo Carletti}, \bibinfo{person}{Pasquale Foggia}, {and} \bibinfo{person}{Mario Vento}.} \bibinfo{year}{2015}\natexlab{}.
\newblock \showarticletitle{VF2 Plus: An improved version of VF2 for biological graphs}. In \bibinfo{booktitle}{\emph{International Workshop on Graph-Based Representations in Pattern Recognition}}. Springer, \bibinfo{pages}{168--177}.
\newblock


\bibitem[Choi and Rhu(2020)]%
        {prema}
\bibfield{author}{\bibinfo{person}{Yujeong Choi} {and} \bibinfo{person}{Minsoo Rhu}.} \bibinfo{year}{2020}\natexlab{}.
\newblock \showarticletitle{Prema: A predictive multi-task scheduling algorithm for preemptible neural processing units}. In \bibinfo{booktitle}{\emph{2020 IEEE International Symposium on High Performance Computer Architecture (HPCA)}}. IEEE, \bibinfo{pages}{220--233}.
\newblock


\bibitem[Cordella et~al\mbox{.}(2004)]%
        {iso_nphard}
\bibfield{author}{\bibinfo{person}{Luigi~P Cordella}, \bibinfo{person}{Pasquale Foggia}, \bibinfo{person}{Carlo Sansone}, {and} \bibinfo{person}{Mario Vento}.} \bibinfo{year}{2004}\natexlab{}.
\newblock \showarticletitle{A (sub) graph isomorphism algorithm for matching large graphs}.
\newblock \bibinfo{journal}{\emph{IEEE transactions on pattern analysis and machine intelligence}} \bibinfo{volume}{26}, \bibinfo{number}{10} (\bibinfo{year}{2004}), \bibinfo{pages}{1367--1372}.
\newblock


\bibitem[Dubey et~al\mbox{.}(2024)]%
        {Llama-3}
\bibfield{author}{\bibinfo{person}{Abhimanyu Dubey}, \bibinfo{person}{Abhinav Jauhri}, \bibinfo{person}{Abhinav Pandey}, \bibinfo{person}{Abhishek Kadian}, \bibinfo{person}{Ahmad Al-Dahle}, \bibinfo{person}{Aiesha Letman}, \bibinfo{person}{Akhil Mathur}, \bibinfo{person}{Alan Schelten}, \bibinfo{person}{Amy Yang}, \bibinfo{person}{Angela Fan}, {et~al\mbox{.}}} \bibinfo{year}{2024}\natexlab{}.
\newblock \showarticletitle{The Llama 3 Herd of Models}.
\newblock \bibinfo{journal}{\emph{CoRR}} (\bibinfo{year}{2024}).
\newblock


\bibitem[Ghodrati et~al\mbox{.}(2020)]%
        {planaria}
\bibfield{author}{\bibinfo{person}{Soroush Ghodrati}, \bibinfo{person}{Byung~Hoon Ahn}, \bibinfo{person}{Joon~Kyung Kim}, \bibinfo{person}{Sean Kinzer}, \bibinfo{person}{Brahmendra~Reddy Yatham}, \bibinfo{person}{Navateja Alla}, \bibinfo{person}{Hardik Sharma}, \bibinfo{person}{Mohammad Alian}, \bibinfo{person}{Eiman Ebrahimi}, \bibinfo{person}{Nam~Sung Kim}, {et~al\mbox{.}}} \bibinfo{year}{2020}\natexlab{}.
\newblock \showarticletitle{Planaria: Dynamic architecture fission for spatial multi-tenant acceleration of deep neural networks}. In \bibinfo{booktitle}{\emph{2020 53rd Annual IEEE/ACM International Symposium on Microarchitecture (MICRO)}}. IEEE, \bibinfo{pages}{681--697}.
\newblock


\bibitem[He et~al\mbox{.}(2016)]%
        {resnet50}
\bibfield{author}{\bibinfo{person}{Kaiming He}, \bibinfo{person}{Xiangyu Zhang}, \bibinfo{person}{Shaoqing Ren}, {and} \bibinfo{person}{Jian Sun}.} \bibinfo{year}{2016}\natexlab{}.
\newblock \showarticletitle{Identity mappings in deep residual networks}. In \bibinfo{booktitle}{\emph{Computer Vision--ECCV 2016: 14th European Conference, Amsterdam, The Netherlands, October 11--14, 2016, Proceedings, Part IV 14}}. Springer, \bibinfo{pages}{630--645}.
\newblock


\bibitem[Jain et~al\mbox{.}(2022)]%
        {PSO}
\bibfield{author}{\bibinfo{person}{Meetu Jain}, \bibinfo{person}{Vibha Saihjpal}, \bibinfo{person}{Narinder Singh}, {and} \bibinfo{person}{Satya~Bir Singh}.} \bibinfo{year}{2022}\natexlab{}.
\newblock \showarticletitle{An overview of variants and advancements of PSO algorithm}.
\newblock \bibinfo{journal}{\emph{Applied Sciences}} \bibinfo{volume}{12}, \bibinfo{number}{17} (\bibinfo{year}{2022}), \bibinfo{pages}{8392}.
\newblock


\bibitem[Kao and Krishna(2022)]%
        {magma}
\bibfield{author}{\bibinfo{person}{Sheng-Chun Kao} {and} \bibinfo{person}{Tushar Krishna}.} \bibinfo{year}{2022}\natexlab{}.
\newblock \showarticletitle{Magma: An optimization framework for mapping multiple dnns on multiple accelerator cores}. In \bibinfo{booktitle}{\emph{2022 IEEE International Symposium on High-Performance Computer Architecture (HPCA)}}. IEEE, \bibinfo{pages}{814--830}.
\newblock


\bibitem[Kim et~al\mbox{.}(2023)]%
        {moca}
\bibfield{author}{\bibinfo{person}{Seah Kim}, \bibinfo{person}{Hasan Genc}, \bibinfo{person}{Vadim~Vadimovich Nikiforov}, \bibinfo{person}{Krste Asanovi{\'c}}, \bibinfo{person}{Borivoje Nikoli{\'c}}, {and} \bibinfo{person}{Yakun~Sophia Shao}.} \bibinfo{year}{2023}\natexlab{}.
\newblock \showarticletitle{Moca: Memory-centric, adaptive execution for multi-tenant deep neural networks}. In \bibinfo{booktitle}{\emph{2023 IEEE International Symposium on High-Performance Computer Architecture (HPCA)}}. IEEE, \bibinfo{pages}{828--841}.
\newblock


\bibitem[Kwon et~al\mbox{.}(2021)]%
        {HDA}
\bibfield{author}{\bibinfo{person}{Hyoukjun Kwon}, \bibinfo{person}{Liangzhen Lai}, \bibinfo{person}{Michael Pellauer}, \bibinfo{person}{Tushar Krishna}, \bibinfo{person}{Yu-Hsin Chen}, {and} \bibinfo{person}{Vikas Chandra}.} \bibinfo{year}{2021}\natexlab{}.
\newblock \showarticletitle{Heterogeneous dataflow accelerators for multi-DNN workloads}. In \bibinfo{booktitle}{\emph{2021 IEEE International Symposium on High-Performance Computer Architecture (HPCA)}}. IEEE, \bibinfo{pages}{71--83}.
\newblock


\bibitem[Li et~al\mbox{.}(2023)]%
        {hasp}
\bibfield{author}{\bibinfo{person}{Hongyi Li}, \bibinfo{person}{Songchen Ma}, \bibinfo{person}{Taoyi Wang}, \bibinfo{person}{Weihao Zhang}, \bibinfo{person}{Guanrui Wang}, \bibinfo{person}{Chenhang Song}, \bibinfo{person}{Huanyu Qu}, \bibinfo{person}{Junfeng Lin}, \bibinfo{person}{Cheng Ma}, \bibinfo{person}{Jing Pei}, {et~al\mbox{.}}} \bibinfo{year}{2023}\natexlab{}.
\newblock \showarticletitle{HASP: Hierarchical asynchronous parallelism for multi-NN tasks}.
\newblock \bibinfo{journal}{\emph{IEEE Trans. Comput.}} \bibinfo{volume}{73}, \bibinfo{number}{2} (\bibinfo{year}{2023}), \bibinfo{pages}{366--379}.
\newblock


\bibitem[Li et~al\mbox{.}(2009)]%
        {mcpat}
\bibfield{author}{\bibinfo{person}{Sheng Li}, \bibinfo{person}{Jung~Ho Ahn}, \bibinfo{person}{Richard~D Strong}, \bibinfo{person}{Jay~B Brockman}, \bibinfo{person}{Dean~M Tullsen}, {and} \bibinfo{person}{Norman~P Jouppi}.} \bibinfo{year}{2009}\natexlab{}.
\newblock \showarticletitle{McPAT: An integrated power, area, and timing modeling framework for multicore and manycore architectures}. In \bibinfo{booktitle}{\emph{Proceedings of the 42nd annual ieee/acm international symposium on microarchitecture}}. \bibinfo{pages}{469--480}.
\newblock


\bibitem[Li et~al\mbox{.}(2011)]%
        {cacti-p}
\bibfield{author}{\bibinfo{person}{Sheng Li}, \bibinfo{person}{Ke Chen}, \bibinfo{person}{Jung~Ho Ahn}, \bibinfo{person}{Jay~B Brockman}, {and} \bibinfo{person}{Norman~P Jouppi}.} \bibinfo{year}{2011}\natexlab{}.
\newblock \showarticletitle{CACTI-P: Architecture-level modeling for SRAM-based structures with advanced leakage reduction techniques}. In \bibinfo{booktitle}{\emph{2011 IEEE/ACM International Conference on Computer-Aided Design (ICCAD)}}. IEEE, \bibinfo{pages}{694--701}.
\newblock


\bibitem[Liu et~al\mbox{.}(2018)]%
        {PNASNet}
\bibfield{author}{\bibinfo{person}{Chenxi Liu}, \bibinfo{person}{Barret Zoph}, \bibinfo{person}{Maxim Neumann}, \bibinfo{person}{Jonathon Shlens}, \bibinfo{person}{Wei Hua}, \bibinfo{person}{Li-Jia Li}, \bibinfo{person}{Li Fei-Fei}, \bibinfo{person}{Alan Yuille}, \bibinfo{person}{Jonathan Huang}, {and} \bibinfo{person}{Kevin Murphy}.} \bibinfo{year}{2018}\natexlab{}.
\newblock \showarticletitle{Progressive neural architecture search}. In \bibinfo{booktitle}{\emph{Proceedings of the European conference on computer vision (ECCV)}}. \bibinfo{pages}{19--34}.
\newblock


\bibitem[Ronneberger et~al\mbox{.}(2015)]%
        {U-net}
\bibfield{author}{\bibinfo{person}{Olaf Ronneberger}, \bibinfo{person}{Philipp Fischer}, {and} \bibinfo{person}{Thomas Brox}.} \bibinfo{year}{2015}\natexlab{}.
\newblock \showarticletitle{U-net: Convolutional networks for biomedical image segmentation}. In \bibinfo{booktitle}{\emph{International Conference on Medical image computing and computer-assisted intervention}}. Springer, \bibinfo{pages}{234--241}.
\newblock


\bibitem[Sandler et~al\mbox{.}(2018)]%
        {Mobilenetv2}
\bibfield{author}{\bibinfo{person}{Mark Sandler}, \bibinfo{person}{Andrew Howard}, \bibinfo{person}{Menglong Zhu}, \bibinfo{person}{Andrey Zhmoginov}, {and} \bibinfo{person}{Liang-Chieh Chen}.} \bibinfo{year}{2018}\natexlab{}.
\newblock \showarticletitle{Mobilenetv2: Inverted residuals and linear bottlenecks}. In \bibinfo{booktitle}{\emph{Proceedings of the IEEE conference on computer vision and pattern recognition}}. \bibinfo{pages}{4510--4520}.
\newblock


\bibitem[Shen et~al\mbox{.}(2023)]%
        {mars}
\bibfield{author}{\bibinfo{person}{Guan Shen}, \bibinfo{person}{Jieru Zhao}, \bibinfo{person}{Zeke Wang}, \bibinfo{person}{Zhe Lin}, \bibinfo{person}{Wenchao Ding}, \bibinfo{person}{Chentao Wu}, \bibinfo{person}{Quan Chen}, {and} \bibinfo{person}{Minyi Guo}.} \bibinfo{year}{2023}\natexlab{}.
\newblock \showarticletitle{MARS: Exploiting multi-level parallelism for DNN workloads on adaptive multi-accelerator systems}. In \bibinfo{booktitle}{\emph{2023 60th ACM/IEEE Design Automation Conference (DAC)}}. IEEE, \bibinfo{pages}{1--6}.
\newblock


\bibitem[Sheridan(2016)]%
        {robotic1}
\bibfield{author}{\bibinfo{person}{Thomas~B Sheridan}.} \bibinfo{year}{2016}\natexlab{}.
\newblock \showarticletitle{Human--robot interaction: status and challenges}.
\newblock \bibinfo{journal}{\emph{Human factors}} \bibinfo{volume}{58}, \bibinfo{number}{4} (\bibinfo{year}{2016}), \bibinfo{pages}{525--532}.
\newblock


\bibitem[Stine et~al\mbox{.}(2007)]%
        {FreePDK45}
\bibfield{author}{\bibinfo{person}{James~E. Stine}, \bibinfo{person}{Ivan Castellanos}, \bibinfo{person}{Michael Wood}, \bibinfo{person}{Jeff Henson}, \bibinfo{person}{Fred Love}, \bibinfo{person}{W.~Rhett Davis}, \bibinfo{person}{Paul~D. Franzon}, \bibinfo{person}{Michael Bucher}, \bibinfo{person}{Sunil Basavarajaiah}, \bibinfo{person}{Julie Oh}, {and} \bibinfo{person}{Ravi Jenkal}.} \bibinfo{year}{2007}\natexlab{}.
\newblock \showarticletitle{FreePDK: An Open-Source Variation-Aware Design Kit}. In \bibinfo{booktitle}{\emph{2007 IEEE International Conference on Microelectronic Systems Education (MSE'07)}}. \bibinfo{pages}{173--174}.
\newblock
\href{https://doi.org/10.1109/MSE.2007.44}{doi:\nolinkurl{10.1109/MSE.2007.44}}


\bibitem[Sun and Luo(2020)]%
        {overview_iso}
\bibfield{author}{\bibinfo{person}{Shixuan Sun} {and} \bibinfo{person}{Qiong Luo}.} \bibinfo{year}{2020}\natexlab{}.
\newblock \showarticletitle{In-memory subgraph matching: An in-depth study}. In \bibinfo{booktitle}{\emph{Proceedings of the 2020 ACM SIGMOD International Conference on Management of Data}}. \bibinfo{pages}{1083--1098}.
\newblock


\bibitem[Tan and Le(2019)]%
        {Efficientnet}
\bibfield{author}{\bibinfo{person}{Mingxing Tan} {and} \bibinfo{person}{Quoc Le}.} \bibinfo{year}{2019}\natexlab{}.
\newblock \showarticletitle{Efficientnet: Rethinking model scaling for convolutional neural networks}. In \bibinfo{booktitle}{\emph{International conference on machine learning}}. PMLR, \bibinfo{pages}{6105--6114}.
\newblock


\bibitem[Tian et~al\mbox{.}(2018)]%
        {autonomous_driving1}
\bibfield{author}{\bibinfo{person}{Yuchi Tian}, \bibinfo{person}{Kexin Pei}, \bibinfo{person}{Suman Jana}, {and} \bibinfo{person}{Baishakhi Ray}.} \bibinfo{year}{2018}\natexlab{}.
\newblock \showarticletitle{Deeptest: Automated testing of deep-neural-network-driven autonomous cars}. In \bibinfo{booktitle}{\emph{Proceedings of the 40th international conference on software engineering}}. \bibinfo{pages}{303--314}.
\newblock


\bibitem[Tran et~al\mbox{.}(2015)]%
        {GpSM}
\bibfield{author}{\bibinfo{person}{Ha-Nguyen Tran}, \bibinfo{person}{Jung-jae Kim}, {and} \bibinfo{person}{Bingsheng He}.} \bibinfo{year}{2015}\natexlab{}.
\newblock \showarticletitle{Fast subgraph matching on large graphs using graphics processors}. In \bibinfo{booktitle}{\emph{International Conference on Database Systems for Advanced Applications}}. Springer, \bibinfo{pages}{299--315}.
\newblock


\bibitem[Ullmann(1976)]%
        {ullmann}
\bibfield{author}{\bibinfo{person}{Julian~R Ullmann}.} \bibinfo{year}{1976}\natexlab{}.
\newblock \showarticletitle{An algorithm for subgraph isomorphism}.
\newblock \bibinfo{journal}{\emph{Journal of the ACM (JACM)}} \bibinfo{volume}{23}, \bibinfo{number}{1} (\bibinfo{year}{1976}), \bibinfo{pages}{31--42}.
\newblock


\bibitem[Wang et~al\mbox{.}(2023)]%
        {CD-MSA}
\bibfield{author}{\bibinfo{person}{Chunyang Wang}, \bibinfo{person}{Yuebin Bai}, {and} \bibinfo{person}{Desen Sun}.} \bibinfo{year}{2023}\natexlab{}.
\newblock \showarticletitle{CD-MSA: cooperative and deadline-aware scheduling for efficient multi-tenancy on DNN accelerators}.
\newblock \bibinfo{journal}{\emph{IEEE Transactions on Parallel and Distributed Systems}} \bibinfo{volume}{34}, \bibinfo{number}{7} (\bibinfo{year}{2023}), \bibinfo{pages}{2091--2106}.
\newblock


\bibitem[Wu et~al\mbox{.}(2019)]%
        {ARVR}
\bibfield{author}{\bibinfo{person}{Carole-Jean Wu}, \bibinfo{person}{David Brooks}, \bibinfo{person}{Kevin Chen}, \bibinfo{person}{Douglas Chen}, \bibinfo{person}{Sy Choudhury}, \bibinfo{person}{Marat Dukhan}, \bibinfo{person}{Kim Hazelwood}, \bibinfo{person}{Eldad Isaac}, \bibinfo{person}{Yangqing Jia}, \bibinfo{person}{Bill Jia}, {et~al\mbox{.}}} \bibinfo{year}{2019}\natexlab{}.
\newblock \showarticletitle{Machine learning at facebook: Understanding inference at the edge}. In \bibinfo{booktitle}{\emph{2019 IEEE international symposium on high performance computer architecture (HPCA)}}. IEEE, \bibinfo{pages}{331--344}.
\newblock


\bibitem[Xia et~al\mbox{.}(2022)]%
        {energy}
\bibfield{author}{\bibinfo{person}{Tian Xia}, \bibinfo{person}{Boran Zhao}, \bibinfo{person}{Jian Ma}, \bibinfo{person}{Gelin Fu}, \bibinfo{person}{Wenzhe Zhao}, \bibinfo{person}{Nanning Zheng}, {and} \bibinfo{person}{Pengju Ren}.} \bibinfo{year}{2022}\natexlab{}.
\newblock \showarticletitle{An energy-and-area-efficient CNN accelerator for universal powers-of-two quantization}.
\newblock \bibinfo{journal}{\emph{IEEE Transactions on Circuits and Systems I: Regular Papers}} \bibinfo{volume}{70}, \bibinfo{number}{3} (\bibinfo{year}{2022}), \bibinfo{pages}{1242--1255}.
\newblock


\bibitem[Yurtsever et~al\mbox{.}(2020)]%
        {automous_driving_unexcepted1}
\bibfield{author}{\bibinfo{person}{Ekim Yurtsever}, \bibinfo{person}{Jacob Lambert}, \bibinfo{person}{Alexander Carballo}, {and} \bibinfo{person}{Kazuya Takeda}.} \bibinfo{year}{2020}\natexlab{}.
\newblock \showarticletitle{A survey of autonomous driving: Common practices and emerging technologies}.
\newblock \bibinfo{journal}{\emph{IEEE access}}  \bibinfo{volume}{8} (\bibinfo{year}{2020}), \bibinfo{pages}{58443--58469}.
\newblock


\bibitem[Zhao et~al\mbox{.}(2022)]%
        {remap}
\bibfield{author}{\bibinfo{person}{Boran Zhao}, \bibinfo{person}{Tian Xia}, \bibinfo{person}{Haiming Zhai}, \bibinfo{person}{Fulun Ma}, \bibinfo{person}{Yan Du}, \bibinfo{person}{Hanzhi Chang}, \bibinfo{person}{Wenzhe Zhao}, {and} \bibinfo{person}{Pengju Ren}.} \bibinfo{year}{2022}\natexlab{}.
\newblock \showarticletitle{REMAP: A spatiotemporal CNN accelerator optimization methodology and toolkit thereof}.
\newblock \bibinfo{journal}{\emph{IEEE Transactions on Computer-Aided Design of Integrated Circuits and Systems}} \bibinfo{volume}{42}, \bibinfo{number}{5} (\bibinfo{year}{2022}), \bibinfo{pages}{1691--1704}.
\newblock


\bibitem[Zhao et~al\mbox{.}(2025)]%
        {isosched}
\bibfield{author}{\bibinfo{person}{Boran Zhao}, \bibinfo{person}{Zihang Yuan}, \bibinfo{person}{Yanbin Hu}, \bibinfo{person}{Haiming Zhai}, \bibinfo{person}{Haoruo Zhang}, \bibinfo{person}{Wenzhe Zhao}, \bibinfo{person}{Tian Xia}, {and} \bibinfo{person}{Pengju Ren}.} \bibinfo{year}{2025}\natexlab{}.
\newblock \showarticletitle{IsoSched: Preemptive Tile Cascaded Scheduling of Multi-DNN via Subgraph Isomorphism}.
\newblock \bibinfo{journal}{\emph{arXiv preprint arXiv:2509.12208}} (\bibinfo{year}{2025}).
\newblock


\bibitem[Zoph et~al\mbox{.}(2018)]%
        {NASNet}
\bibfield{author}{\bibinfo{person}{Barret Zoph}, \bibinfo{person}{Vijay Vasudevan}, \bibinfo{person}{Jonathon Shlens}, {and} \bibinfo{person}{Quoc~V Le}.} \bibinfo{year}{2018}\natexlab{}.
\newblock \showarticletitle{Learning transferable architectures for scalable image recognition}. In \bibinfo{booktitle}{\emph{Proceedings of the IEEE conference on computer vision and pattern recognition}}. \bibinfo{pages}{8697--8710}.
\newblock


\end{thebibliography}










\end{document}